\documentclass[11pt]{article}
\usepackage{amsmath}
\usepackage{amsfonts}
\usepackage{amssymb}
\usepackage{graphicx}
\usepackage{subfigure}
\usepackage{epsfig}

\def\bea{\begin{eqnarray}}
\def\eea{\end{eqnarray}}
\begin{document}
\begin{center}
\LARGE {\bf Mixing and Decoherence in Continuous-time quantum walks
on long-range interacting cycles
 }
\end{center}
\begin{center}
{\bf S. Salimi {\footnote {E-mail: shsalimi@uok.ac.ir}}, R. Radgohar {\footnote {E-mail: r.radgohar@uok.ac.ir}}}\\
 {\it Faculty of Science,  Department of Physics, University of Kurdistan, Pasdaran Ave., Sanandaj, Iran} \\
 \end{center}
\vskip 3cm
\begin{center}
{\bf{Abstract}}
\end{center}
 We study the effect of small decoherence in continuous-time quantum walks
on long-range interacting cycles (LRICs), which are constructed by
connecting all the two nodes of distance $m$ on the cycle graph. In
our investigation, each node is continuously monitored by an
individual point contact (PC), which induces the decoherence
process. We obtain the analytical probability distribution and the
mixing time upper bound. Our results show that, for small rates of
decoherence, the mixing time upper bound is independent of distance
parameter $m$ and is proportional to inverse of decoherence rate.

\newpage

\section{Introduction}

 Random walks on graphs have broad applications in various fields of
 mathematics, computer science and natural sciences, such as
 mathematical modeling of physical systems and simulated
 annealing~\cite{AKV}. The quantum mechanical analog of the
 random walks on complex networks has been studied with
 respect to the localization and delocalization transition in the
 presence of site disorder~\cite{KHC, XP,KWRD}. Quantum walks (QWs) have been largely divided into two standard
 variants, the discrete-time QWs (DTQWs)~\cite{DJ, TBCA} and the
 continuous-time QWs (CTQWs)~\cite{ABT}. The DTQWs have been investigated on trees~\cite{CHKS1},
 on random environments~\cite{NKRE}, for single and entangled particles~\cite{CMCH} and also
in~\cite{AKV,CMRSRL, ADGB}. In the recent years, the CTQWs have
been studied on
 $\emph{n}$-cube~\cite{MR}, star graph~\cite{ssa1, xp1}, small-world
network~\cite{MPB}, quotient graph~\cite{SS}, line \cite{jas1,jas3,
konno}, dendrimer~\cite{OMV}, distance regular graph~\cite{jas2},
circulant  Bunkbeds~\cite{LRS}, odd graph
 \cite{shs} and on decision tree~\cite{FG, konno2}.

  In all of these  cited works have been supposed that we have a closed quantum system without any
 interaction with its environment. Firstly, Kendon and Tregenna considered the effect of decoherence in
 quantum walks in ''Decoherence can be useful in quantum walks''~\cite{KT}. By numerical observation, they found that
 a small amount of decoherence can be useful to decrease the mixing
 time of discrete quantum walks on cycles. Then deoherence in quantum walks has been
 studied on line~\cite{RSA}, on circulant~\cite{LT} and on
 hypercube~\cite{WS}. In '' Mixing and
 decoherence in continuous-time quantum walks on Cycles"~\cite{FST},
 it has been provided an analytical counterpart to Kendon and Tregenna's
 result for the continuous-time quantum walk on cycles. Its results
 show that, for small rates of decoherense, the mixing time decreases
 linearly with decoherence while for large rates of decoherence,
 the mixing time increases linearly toward the classical limit.
 Moreover, for the middle region of decoherence rates, the numerical data
 confirms the existence of a unique optimal rate in which the
 mixing time is minimal.

  In this paper, we consider the effect of
 decoherence in continuous-time quantum walks on long-range
 interacting cycles (LRICs) as the extensions of the cycle graphs. LRICs are constructed by connecting all
 the two nodes of distance $m$ on the cycle graph (nearest-neighboring
 lattice). A detailed description of the network structure
 will be given in the next section. Numerical analysis of CTQWs on LRICs
 has been provided in ''Coherent exciton transport and trapping on long-range interacting
 cycles''~\cite{XP}. We take a small amount of decoherence into
 account and by analytical technique show it can decrease the mixing time in
continuous-time quantum walks on \clearpage
\begin{figure}[h]
\includegraphics{rr.eps}
\vspace{.85cm} \caption{Long-range interaction cycles $G(8; 3)$ and
$G(10; 4)$}\label{a}
\end{figure}

LRICs. For this end, we use of Gurvitz model~\cite{AG}. In this
model, each of vertex is monitored by a corresponding point contact
induced the decoherence process. We calculate the analytical
probability distribution then obtain the mixing time upper bound for
small rates of decoherence. We show that it is independent of the
distance parameter $m$ and is proportional to inverse of decoherence
rate. Our paper is structured as follows: After a detailed
description of the network structure LRICs in Sec. 2, we study
continuous-time quantum walks over the underlying structures in Sec.
3. In Sec. 4 we consider the effect of decoherence in CTQWs on LRICs
and in Sec. 5 we focus on the decoherent CTQWs when the decoherent
rate is small. In Sec. 6, we define the mixing time and obtain its
upper bound. Finally, in Sec. 7 the conclusions are presented.

\section{Structure of LRICs}
To construct long-range interacting cycles (LRICs), we can use of
the following rule: First, we construct a cycle graph of $N$ nodes
where each node connected to its two nearest neighbor nodes. Second,
all the two nodes of distance $m$ on the cycle graph are connected
by additional bonds. LRICs are denoted by $G(N,m)$ that $N$ is the
network size and $m$ is long-range interaction parameter. LRIC is a
one-dimensional lattice with periodic boundary conditions and all
nodes of the networks have four bonds~\cite{XP}. The structures of
$G(8,3)$ and $G(10,4)$ are illustrated in Fig. 1.

\section{CTQWs on LRICs}
CTQWs on LRICs is obtained by replacing the Hamiltonian of the
system by the classical transfer matrix, i.e., $H=-T$~\cite{FG}. The
transfer matrix $\textbf{T}$ relates to the Laplace matrix by
$T=-\gamma A$, where for simplicity we assume the transmission rates
$\gamma$ of all bonds to be equal and set $\gamma=1$ in the
following. In Laplace matrix $\textbf{A}$, nondiagonal elements
$A_{ij}$ equal to 1 if nodes $i$ and $j$ be connected and 0
otherwise. The diagonal elements $\textbf{A}_{ii}$ follow as
$\textbf{A}_{ii}=-k_{i}$ that $k_{i}$ is the degree of vertex $i$.
The basis vectors $|j\rangle$ associated with the nodes $j$ of the
graph span the whole accessible Hilbert space.\clearpage
 Then, the Hamiltonian matrix $H$ of $G(N,m)$
$(m\geq2)$ takes the following form,
\begin{eqnarray}\label{1}
   H_{ij}=\langle i|H|j \rangle=\left\{
                                \begin{array}{ll}
                                  -4, & \hbox{if i= j;} \\
                                  1, & \hbox{if $i=j\pm 1$;} \\
                                  1, & \hbox{if $i=j\pm m$;} \\
                                  0, & \hbox{Otherwise.}
                                \end{array}
                              \right.
        \end{eqnarray}

The Hamiltonian acting on the state $|j\rangle$ can be written as

\begin{eqnarray}\label{2}
   H|j\rangle=-4|j\rangle+|j-1\rangle+|j+1\rangle+|j-m\rangle+|j+m\rangle,
   \end{eqnarray}

which is the discrete version of the Hamiltonian for a free particle
moving on a lattice. It is well known in solid state physics that
the solutions of the Schr\"{o}dinger equation for a particle moving
freely in regular potential are Bloch function~\cite{MZ, CK}. Thus,
the time independent Schr\"{o}dinger equation is given by

\begin{eqnarray}\label{3}
   \textbf{H}|\Phi_{\theta}\rangle=E_{\theta}|\Phi_{\theta}\rangle,
   \end{eqnarray}

where the eigenstates $|\Phi_{\theta}\rangle$ are Bloch states and
can be written as a linear combination of states $|j\rangle$
localized at nodes $j$,

\begin{eqnarray}\label{4}
   |\Phi_{\theta}\rangle=\frac{1}{\sqrt{N}}\sum_{j=1}^{N}e^{-i\theta j}|j\rangle.
   \end{eqnarray}
The projection on the state $|j\rangle$ is $\Phi_{\theta}(j)=\langle
j|\Phi_{\theta}\rangle= \frac{1}{\sqrt{N}} e^{-i\theta j}$, which is
nothing but the Bloch relation $\Phi_{\theta}(j+1)=
e^{-i\theta}\Phi_{\theta}(j)$~\cite{MZ, CK}.
\\Now the energy is obtained from Eqs. (3) and (4) as
\begin{eqnarray}\label{5}
  E_{\theta}= -4+2\cos(\theta)+2\cos(m\theta),
   \end{eqnarray}
for $j=0, 1, ..., N-1$. The classical and quantum transition
probabilities between two nodes can be written as

\begin{eqnarray}\label{6}
   P_{k,j}(t)=\sum_{\theta}e^{-tE_{\theta}}\langle k|\Phi_{\theta}\rangle\langle
   \Phi_{\theta}|j\rangle,
   \end{eqnarray}

\begin{eqnarray}\label{7}
   \pi_{k,j}(t)=|\alpha_{k,j}(t)|^{2}=|\sum_{\theta}e^{-itE_{\theta}}\langle k|\Phi_{\theta}\rangle\langle \Phi_{\theta}|j\rangle|^{2}.
   \end{eqnarray}
\clearpage
\begin{figure}[htb]
\vspace{5cm} \includegraphics{double.eps} \vspace{.00001cm} \caption{Fig. 2(a) shows
point contact detector $j$ monitoring the electron in dot $j$ and
Fig. 2(b) shows point contact $j$ in the presence of the electron in
dot $j+1$.}\label{a}
\end{figure}

\section{The Decoherent CTQWs on LRICs}
\textbf{Gurvitz's model}
\\To analyze the decoherent continuous-time quantum walks on LRICs,
we use analytical model developed by Gurvitz~\cite{AG, GFM}.
\\In this model, every node is regarded as a quantum dot. Thus LRIC
is represented by a set of the identical tunnel-coupled quantum dots
(QDs). The walks are done by an electron initially placed in one of
dots. A ballistic point-contact (PC) is placed near every dot that
is taken as a noninvasive detector. We assume that all point
contacts are identical. Also, they are placed far enough from
quantum dots so that the tunneling between them is negligible.
Moreover, for simplicity, we consider electrons as spinless
fermions~\cite{FST}. Each PC continuously monitors the attached
quantum dot. This measurement process induces decoherence to
electron walks as is shown in Fig. 2. Firstly, we study the simple
quantum walks with the Hamiltonian~\cite{APS}
\begin{eqnarray}\label{8}
\begin{array}{cc}
  H_{0}= & -\displaystyle\sum_{ij}\Delta_{ij}(t)(\hat{c}^{\dag}_{i}\hat{c}_{j}+\hat{c}_{i}\hat{c}^{\dag}_{j})+\displaystyle\sum_{j}\epsilon_{j}(t)\hat{c}^{\dag}_{j}\hat{c}_{j}\hspace{2cm}\\
   &  \\
   \equiv\hspace{-.4cm}&-\displaystyle\sum_{ij}\Delta_{ij}(t)(|i\rangle\langle j|+|j\rangle\langle
 i\rangle)+\displaystyle\sum_{j}\epsilon_{j}(t)|j\rangle\langle j|,\hspace{1.5cm}
\end{array}
   \end{eqnarray}

where $|j\rangle=\hat{c}^{\dag}_{j}|0\rangle$ denotes the state that
the electron is placed at dot $j$. $\Delta_{ij}$ is the hopping
amplitude between dots $i,j$ and $\epsilon_{j}(t)$ is the on-site
dot energy. We assume the constant hopping amplitude between linked
dots and no on-site terms. For simplicity, we renormalize the time,
so that it becomes dimensionless~\cite{SF}. Hence, the Hamiltonian
has the form

\begin{eqnarray}\label{9}
 H_{0}=\frac{1}{4}\sum_{j=0}^{N-1}(\hat{c}^{\dag}_{j+1}\hat{c}_{j}+\hat{c}_{j}^{\dag}\hat{c}_{j+1}+\hat{c}^{\dag}_{j+m}\hat{c}_{j}+\hat{c}_{j}^{\dag}\hat{c}_{j+m}).
   \end{eqnarray}

The tunneling Hamiltonian $H_{PC,j}$ describing electron transport
in the point contact $j$ can be written as~\cite{SF, AG}

\begin{eqnarray}\label{10}
H_{PC,j}=\sum_{l}E_{l,j}\hat{a}^{\dag}_{l,j}\hat{a}_{l,j}+\sum_{r}E_{r,j}\hat{a}^{\dag}_{r,j}\hat{a}_{r,j}+\sum_{lr}\Omega_{lr,j}(\hat{a}^{\dag}_{r,j}\hat{a}_{l,j}+H.C.).
   \end{eqnarray}
Here, $\hat{a}^{\dag}_{l,j}(\hat{a}_{l,j})$ and
$\hat{a}^{\dag}_{r,j}(\hat{a}_{r,j})$ are the creation
(annihilation) operators in the left and right reservoirs of
detector $j$, respectively, and $\Omega_{lr,j}$ is the hopping
amplitude between the states $l$ and $r$ of detector $j$.
\\
\\The presence of an electron in the left dot results in an effective
increase of the point-contact barrier
$(\Omega_{lr}\rightarrow\Omega_{l,r}+\delta\Omega_{l,r})$, we can
represent the interaction term as~\cite{SF, AG}

\begin{eqnarray}\label{11}
 H_{int,j}=\sum_{l,r}\delta\Omega_{lr,j}\hat{c}^{\dag}_{j}\hat{c}_{j}(\hat{a}^{\dag}_{l,j}\hat{a}_{r,j}+H.C.).
   \end{eqnarray}

Thus, the entire system Hamiltonian can be described by

\begin{eqnarray}\label{12}
 H=H_{0}+\sum_{j=0}^{N-1}H_{PC,j}+H_{int,j}.
   \end{eqnarray}

We suppose for simplicity that the hopping amplitude of $j$-th
point-contact is weakly dependent on the states $l$ and $r$, so that
it can be replaced by its average value,
$(\Omega_{lr,j}\simeq\bar{{\Omega}})$ and
$\delta\Omega_{lr,j}\simeq\delta\bar{\Omega}$. The occupation of the
quantum dot can be measured through the variation of the detector
current $\Delta I=I_{2}-I_{1}$ where $I_{1}=e2\pi
\bar{\Omega}^{2}\rho_{l,j}\rho_{r,j}V_{j}$ is the detector current
when the electron occupies the first dot, Fig. 2(a) and $I_{2}=e2\pi
(\bar{\Omega}^{2}+\delta\bar{\Omega})^{2}\rho_{l,j}\rho_{r,j}V_{j}$
is the current flowing through the detector in the presence of the
electron in the second dot, Fig. 2(b). The density of states in the
left and right reservoirs of detector $j$ are $\rho_{l,j}$ and
$\rho_{r,j}$ and the voltage bias is the variation of the chemical
potentials in the left and right reservoirs in detector i.e.
$V_{j}=\mu_{l,j}-\mu_{r,j}$~\cite{GFM}. Now, we are ready to write
Schr\"{o}dinger equation for the entire system with Hamiltonian $H$.
The effect of the detector on the quantum dot can be obtained by
tracing out the detector states. Gurvitz has shown, for a double-dot
and detector together (Fig. 2), Schr\"{o}dinger equation results in
the evolution of reduced density matrix traced over all states of
detector which coincides with the Bloch-type rate
equations~\cite{AG, GFM}. These equations are as

\begin{eqnarray}\label{13}
\begin{array}{c}
  \dot{\rho}_{j,j}=i\Omega_{0}(\rho_{j,j+1}-\rho_{j+1,j}), \\
   \\
   \dot{\rho}_{j+1,j+1}=i\Omega_{0}(\rho_{j+1,j}-\rho_{j,j+1}),\\
  \\
\dot{\rho}_{j,j+1}=i\epsilon_{j}\rho_{j,j+1}+i\Omega_{0}(\rho_{j,j}-\rho_{j+1,j+1})-\frac{\Gamma}{2}\rho_{j,j+1},
\end{array}
   \end{eqnarray}

where $\epsilon_{j}=E_{j}-E_{j+1}$ and $\Omega_{0}$ is the coupling
between the left and right dots. Also,
${\rho}_{j,j+1}(t)={\rho}^{\ast}_{j+1,j}(t)$ are the off-diagonal
reduced density matrix elements and the diagonal terms of this
density matrix ${\rho}_{j,j}(t), {\rho}_{j+1,j+1}(t)$ are the
probabilities of finding the electron in $j$th dot and in $j+1$th
dot, respectively. Moreover,
$\Gamma=(\sqrt{\frac{I_{1}}{e}}-\sqrt{\frac{I_{2}}{e}})^{2}\frac{V_{j}}{2\pi}$
is decoherence rate due to continuous observation with a
non-invasive detector~\cite{AG, GFM}. Applying this model to our
system results in

  \begin{eqnarray}\label{14}
\begin{array}{cc}
  \frac{d}{dt}\rho_{j,k}(t)= & \frac{i}{4}[-\rho_{j-1,k}-\rho_{j+1,k}-\rho_{j-m,k}-\rho_{j+m,k}+\rho_{j,k-1}  \\
   &  \\
   & +\rho_{j,k+1}+\rho_{j,k-m}+\rho_{j,k+m}]-\Gamma(1-\delta_{j,k})\rho_{j,k}
\end{array}
  \end{eqnarray}
Our subsequent analysis will focus on the variable $S_{j,k}$ defined as~\cite{FST}
\begin{eqnarray}\label{15}
 S_{j,k}=i^{k-j}\rho_{j,k}
   \end{eqnarray}
Substituting Eq. (15) into Eq. (14), we have
 \begin{eqnarray}\label{16}
\begin{array}{cc}
  \frac{d}{dt}S_{j,k}=  & \frac{1}{4}[-S_{j-1,k}+S_{j+1,k}-i^{-m+1}S_{j-m,k}-i^{m+1}S_{j+m,k}-S_{j,k-1} \\
   &\\
 & +S_{j,k+1}+i^{m+1}S_{j,k-m}+i^{-m+1}S_{j,k+m}]-\Gamma(1-\delta_{j,k})S_{j,k}
\end{array}
    \end{eqnarray}

\section{Small Decoherence}
We consider the coherent continuous-time quantum walks when $\Gamma
N\ll1$. Eq. (16) can rewrite as the perturbed linear operator
equation~\cite{FST}
\begin{eqnarray}\label{17}
 \frac{d}{dt}S(t)=(L+U)S(t),
   \end{eqnarray}
where the linear operators $L$ and $U$ are
\begin{eqnarray}\label{18}
  \begin{array}{cc}
    L_{(\alpha,\beta)}^{(\mu,\nu)}= & \frac{1}{4}[-\delta_{\alpha,\mu+1}\delta_{\beta,\nu}+\delta_{\alpha,\mu-1}\delta_{\beta,\nu}-i^{-m+1}\delta_{\alpha,\mu+m}\delta_{\beta,\nu}-i^{m+1}\delta_{\alpha,\mu-m}\delta_{\beta,\nu} \\
     &  \\
     & -\delta_{\alpha,\mu}\delta_{\beta,\nu+1}+\delta_{\alpha,\mu}\delta_{\beta,\nu-1}+i^{m+1}\delta_{\alpha,\mu}\delta_{\beta,\nu+m}+i^{-m+1}\delta_{\alpha,\mu}\delta_{\beta,\nu-m}]
  \end{array}
  \end{eqnarray}

\begin{eqnarray}\label{19}
U_{(\alpha,\beta)}^{(\mu,\nu)}=-\Gamma(1-\delta_{\alpha,\beta})\delta_{\alpha,\mu}\delta_{\beta,\nu},
   \end{eqnarray}
that $L$ is a $N^{2}\times N^{2}$ matrix and
$L^{(\mu,\nu)}_{(\alpha,\beta)}$ is the entry of $L$ indexed by the
row index $(\mu,\nu)$ and the column index $(\alpha,\beta)$. Also,
$U$ has the same behavior. By the above Substituting, we obtain
\begin{eqnarray}\label{20}
 \frac{d}{dt}S_{\alpha,\beta}=\sum_{\mu,\nu=0}^{N-1}(L_{(\alpha,\beta)}^{(\mu,\nu)}+U_{(\alpha,\beta)}^{(\mu,\nu)})S_{\mu,\nu},
   \end{eqnarray}

where $0\leq\alpha,\beta,\mu,\nu\leq N-1$.  The initial conditions are
 \begin{eqnarray}\label{21}
  \rho_{\alpha,\beta}(0)=S_{\alpha,\beta}(0)=\delta_{\alpha,0}\delta_{\beta,0}.
    \end{eqnarray}

To obtain zero-order solution of Eq. (20), we require to an
expansion on the eigenvectors of $L_{(\alpha,\beta)}^{(\mu,\nu)}$ or
\begin{eqnarray}\label{22}
 \sum_{\mu,\nu=0}^{N-1}L_{(\alpha,\beta)}^{(\mu,\nu)}V_{(\mu,\nu)}^{(k,l)}=\lambda_{(k,l)}^{0}V_{(\alpha,\beta)}^{(k,l)},
   \end{eqnarray}
where $0\leq k,l\leq N-1$. After some algebra with Eq. (22), one can
obtain
\begin{eqnarray}\label{23}
\lambda_{k,l}=i[\sin(\frac{\pi(k+l)}{N})\cos(\frac{\pi(k-l)}{N})]+i^{m}[\sin(\frac{\pi
m(k+l)}{N})\cos(\frac{\pi m(k-l)}{N})]
   \end{eqnarray}
and
\begin{eqnarray}\label{24}
\lambda_{k,l}=i[\sin(\frac{\pi(k+l)}{N})\cos(\frac{\pi(k-l)}{N})]+i^{m+1}[\sin(\frac{\pi
m(k+l)}{N})\sin(\frac{\pi m(k-l)}{N})],
   \end{eqnarray}
when $m$ is an odd and even number, respectively. Eigenvectors of
$L$ are given by
\begin{eqnarray}\label{25}
 V_{(\mu,\nu)}^{(k,l)}=\frac{1}{N}\exp(\frac{2\pi i}{N}(k\mu+l\nu)).
   \end{eqnarray}

Now, we consider eigenvalues of the unperturbed linear operator $L$
(Eqs. (23), (24)) carefully. We investigate the important
degeneracies of the eigenvalues $\lambda_{k,l}$ of $L$ that lead to
non-zero off-diagonal contribution of $U$. Firstly, due to symmetry
of Eq. (23), we have $\lambda_{k,l}=\lambda_{l,k}$, while their
eigenvectors (Eq. (25)) are clearly different (Note that in Eq. (24)
there is not such symmetry). The second subset of degenerate
eigenvalues appear when we replace $k+l\equiv0 (mod N)$ in Eqs. (23)
, (24). one can see the corresponding eigenvectors are not the same.
The third subset of the degenerate eigenvalues reveal when we set
$k=l$.
\\First-order correction terms are given
by the diagonal elements of Eq. (19) calculated on eigenvectors of
Eq. (25). For the first subset of degenerate eigenvalues, they are
equal to $-\Gamma\frac{(N-2)}{N}$. By introducing this eigenvalue
perturbation to each pair of $\lambda_{(k,l)}$ with $k\neq l$ and
$k+l\neq N$, the degeneracy of the first subset is removed. The
degeneracy of the second subset is absent in our case since the
their eigenvectors are anyway excluded from the final solution by
initial condition~\cite{SF}. In our discursion, there is not the
degeneracy of the third subset since $U$ is diagonal over the
corresponding eigenvectors~\cite{FST}. For these eigenvalues, the
correction terms are $-\Gamma\frac{(N-1)}{N}$. Thus, as mentioned
in~\cite{FST, SF}, Eq. (20) can be expressed in terms of
eigenvectors of Eq.(25)
\begin{eqnarray}\label{26}
 S_{\alpha,\beta}(0)=\frac{\delta_{\alpha,\beta}}{N}+\frac{1}{N^{2}}\sum_{k,l=0}^{N-1}(1-\delta_{k+l,0}-\delta_{k+l,N})\exp[\frac{2\pi i(k\alpha+l\beta)}{N}]
   \end{eqnarray}

The full solution is of the form
\begin{eqnarray}\label{27}
\begin{array}{cc}
  S_{\alpha,\beta}(t)= & \frac{\delta_{\alpha,\beta}}{N}+\displaystyle\sum_{k,l=0}^{N-1}\frac{1-\delta_{k+l,0}-\delta_{k+l,N}}{N^{2}}\exp[\frac{2\pi i(k\alpha+l\beta)}{N}] e^{t(\lambda_{(k,l)}+\tilde{\lambda}_{(k,l)})}\\
   \end{array}
   \end{eqnarray}
The probability distribution of the continuous-time quantum walks is
given by the diagonal elements of the reduced density matrix
$P_{j}(t)=S_{j,j}(t)$, that is

\begin{eqnarray}\label{28}
\begin{array}{cc}
  P_{j}(t)= & \frac{1}{N}+\displaystyle\sum_{k,l=0}^{N-1}\frac{1-\delta_{k+l,0}-\delta_{k+l,N}}{N^{2}}\exp[\frac{2\pi i(k+l)j}{N}] e^{t(\lambda_{(k,l)}+\tilde{\lambda}_{(k,l)})}\\
   \end{array}
   \end{eqnarray}

 Now, we want to rewritten the above equation for odd and even values of $m$, respectively.

\textbf{odd $m$}

The solution is of Eq. (20) is

\begin{eqnarray}\label{29}
\begin{array}{cc}
  P_{j}(t)= & \frac{1}{N}+\displaystyle\sum_{k,l=0}^{N-1}\frac{1-\delta_{k+l,0}-\delta_{k+l,N}}{N^{2}}[\delta_{k,l}e^{-\Gamma\frac{N-1}{N}t}+(1-\delta_{k,l})e^{-\Gamma\frac{N-2}{N}t}]e^{\frac{2\pi i(k+l)j}{N}}\hspace{1cm} \\
   &  \\
  &
  \times\exp[it\sin(\frac{\pi(k+l)}{N})\cos(\frac{\pi(k-l)}{N})+i^{m}t\sin(\frac{\pi m(k+l)}{N})\sin(\frac{\pi m(k-l)}{N})].\hspace{1.5cm}
\end{array}
   \end{eqnarray}
for odd $m$.
\\
\textbf{even $m$}

The full solution of Eq. (20) is

\begin{eqnarray}\label{30}
\begin{array}{cc}
  P_{j}(t)= & \frac{1}{N}+\displaystyle\sum_{k,l=0}^{N-1}\frac{1-\delta_{k+l,0}-\delta_{k+l,N}}{N^{2}}\times e^{-\Gamma\frac{N-1}{N}t}\times\exp[\frac{2\pi
i(k+l)j}{N}]\hspace{1cm} \\
   &  \\
  &
  \times\exp[it\sin(\frac{\pi(k+l)}{N})\cos(\frac{\pi(k-l)}{N})+i^{m+1}t\sin(\frac{\pi
m(k+l)}{N})\sin(\frac{\pi m(k-l)}{N})].
\end{array}
   \end{eqnarray}
for even $m$.

\section{Mixing time}
To define the mixing time of continuous-time quantum walks, we use
the principal motivation to studying random walks. In computer
science, the probabilistic algorithm provide the best solution for
many problems. Thus the precise solution is obtained by a
well-chosen sampling distribution. Generating the such distribution
is a matter of mapping the uniform distribution into the desired
one. Hence, it is important to get a truly uniform
distribution~\cite{MDAHMSS}. The mixing time $T_{mix}(\epsilon)$ is
defined as the number of steps needed before the distribution is
guaranteed to be $\epsilon$-close to the uniform distribution, or
\begin{eqnarray}\label{31}
T_{mix}(\epsilon)=\min\{T:
\sum_{j=0}^{N-1}|P_{j}(t)-\frac{1}{N}|\leq\epsilon\},
   \end{eqnarray}

that $P_{j}(t)$ is the probability distribution of quantum walk on
node $j$ of graph $G$ and $\frac{1}{N}$ is the uniform distribution
over graph $G$.
\\Based on the above analysis, we will be interested in decreasing the mixing time.
\\Firstly, we want to calculate the upper bound on the
$\epsilon$-uniform mixing time $T_{mix}(\epsilon)$ of Eq. (29)(for
odd $m$).
\\We define

\begin{eqnarray}\label{32}
M_{j}(t)=\frac{1}{N}\sum_{k=0}^{N-1}\exp[it\sin(\frac{2\pi
k}{N})+i^{m}t\sin(\frac{2\pi km}{N})]e^{\frac{2\pi ikj}{N}}.
   \end{eqnarray}

Hence, we have
\\$M_{j}^{2}(\frac{t}{2})=\frac{1}{N^{2}}\displaystyle\sum_{k,l=0}^{N-1}e^{t\lambda_{(k,l)}}e^{\frac{2\pi
i(k+l)j}{N}}$ ,
$M_{2j}(t)=\frac{1}{N}\displaystyle\sum_{k=0}^{N-1}e^{t\lambda_{(k,k)}}e^{\frac{2\pi
ik(2j)}{N}}$.

Hence, Eq. (29) can be rewritten as

\begin{eqnarray}\label{33}
|P_{j}(t)-\frac{1}{N}|=e^{-\Gamma\frac{N-2}{N}t}|M_{j}^{2}(\frac{t}{2})-\frac{1}{N}+\frac{e^{-\frac{\Gamma
t}{N}}-1}{N}[M_{2j}(t)-\frac{2-N mod 2}{N}]|.
   \end{eqnarray}

Since $|M_{j}(t)|\leq1$, we obtain

\begin{eqnarray}\label{34}
\begin{array}{cc}
  |P_{j}(t)-\frac{1}{N}|\leq & e^{-\Gamma\frac{N-2}{N}t}|1-\frac{1}{N}+\frac{e^{-\frac{\Gamma
t}{N}}-1}{N}[1-\frac{2-N mod 2}{N}]|. \\
  &  \\
   \leq\hspace{-1.8cm} & e^{-\Gamma\frac{N-2}{N}t}|1+\frac{e^{-\frac{\Gamma
   t}{N}}-1}{N}(1-\frac{2}{N})|.\hspace{1.8cm}
\end{array}
   \end{eqnarray}

Based on the definition of time mixing,

\begin{eqnarray}\label{35}
\sum_{j=0}^{N-1}|P_{j}(t)-\frac{1}{N}|\leq
e^{-\Gamma\frac{N-2}{N}t}(N+e^{\frac{-t\Gamma}{N}}-1)
   \end{eqnarray}

Because of $e^{-\frac{\Gamma t}{N}}\leq 1$ , the above equation
reduces to $Ne^{-\Gamma\frac{N-2}{N}t}\leq\epsilon$.
\\As a result we obtain the mixing time upper bound of
\begin{eqnarray}\label{36}
T_{mix}(\epsilon)<\frac{1}{\Gamma}\ln(\frac{N}{\epsilon})[1+\frac{2}{N-2}].
   \end{eqnarray}
This relation is in agreement with the result is mentioned
in~\cite{FST} for cycle.
\\Note that in this case, the mixing time lower bound can not derive of Eq.
(33) easily, since there is a relation between
$M_{j}^{2}(\frac{t}{2})$ and $M_{2j}(t)$.
\\
Now, we calculate the upper bound on $T_{mix}(\epsilon)$ of Eq.
(30)(for even $m$). \\We define

\begin{eqnarray}\label{37}
\begin{array}{cc}
  M_{j}(t)= & \frac{1}{N^{2}}\displaystyle\sum_{k,l=0}^{N-1}\exp[i\frac{t}{2}(\sin(\frac{2\pi
k}{N})+\sin(\frac{2\pi l}{N}))]e^{\frac{2\pi i(k+l)
j}{N}} \\
   &  \\
   &\times\exp[-i^{m+1}\frac{t}{2}(\cos(\frac{2\pi km}{N})
-\cos(\frac{2\pi l m}{N}))].\hspace{1.3cm}
\end{array}
   \end{eqnarray}
Thus, we have
\begin{eqnarray}\label{38}
\begin{array}{cc}
   |P_{j}(t)-\frac{1}{N}|=& e^{-\Gamma\frac{N-1}{N}t}|M_{j}(t)-\frac{1}{N}| \\
   & \\
   \leq\hspace{-1.8cm}&  e^{-\Gamma\frac{N-1}{N}t}[1+\frac{1}{N}]\hspace{1cm}
\end{array}
   \end{eqnarray}

and with summation over $j$ and using the mixing time definition, we
get $Ne^{-\Gamma\frac{N-1}{N}t}\leq\epsilon$.

This gives the mixing time upper bound of
\begin{eqnarray}\label{39}
T_{mix}(\epsilon)\leq\frac{1}{\Gamma}[1+\frac{1}{N-1}]\ln(\frac{N}{\epsilon
}).
   \end{eqnarray}

The lower bound of mixing time can derived by the first equality
of Eq. (38), as
\begin{eqnarray}\label{40}
\begin{array}{cc}
   |P_{j}(t)-\frac{1}{N}|=& e^{-\Gamma\frac{N-1}{N}t}|M_{j}(t)-\frac{1}{N}| \\
   & \\
   =0\hspace{-2.3cm}
\end{array}
   \end{eqnarray}
where in the last inequality, we set $M_{j}(t)=\frac{1}{N}$. In
other words, in this time the quantum walk completely reaches to
uniform distribution $\frac{1}{N}$.

Thus, the lower bound of mixing time follows as

\begin{eqnarray}\label{41}
T_{mix}(\epsilon)\geq0.
   \end{eqnarray}

Eqs. (36) and (39) show that the $T_{mix}$ upper bound is
independent of the distance parameter $m$. Moreover, since we
approximated the coefficient of exponentially function in Eqs. (34),
(38), the mixing time upper bound is exactly proportional to
$\frac{1}{\Gamma}$ that accord with Fedichkin, Solenov and Tamon 's
result for the continuous-time quantum walks on cycles~\cite{FST}.
Also, these relations prove that the mixing time upper bound for
even $m$ is smaller than mixing time upper bound for odd $m$.
\section{Conclusion}
We have studied continuous-time quantum walks on long-range
interacting cycles (LRICs) under small-decoherence $\Gamma N\ll
1$. We obtained the probability distribution analytically and
found that the mixing time upper bound for odd values of $m$
($T_{mix}(\epsilon)<\frac{1}{\Gamma}\ln(\frac{N}{\epsilon})[\frac{N}{N-2}]$)
is larger than the mixing time upper bound for even $m$
($T_{mix}(\epsilon)\leq\frac{1}{\Gamma}\ln(\frac{N}{\epsilon
})[\frac{N}{N-1}]$). These relations show that the $T_{mix}$ upper
bound is inversely proportional to decoherence rate $\Gamma$ and
is independent of the distance parameter $m$. Also, we proved that
for even $m$ the lower bound time mixing is zero. In other words,
we have shown that $T_{mix}(\epsilon)$ decrease with $\Gamma$ at
least as fast as  $1/\Gamma$.

\end{document}